\documentclass[12pt]{article}

\usepackage[utf8]{inputenc}
\usepackage{times}

\usepackage{graphicx}% Include figure files
\usepackage{dcolumn}% Align table columns on decimal point
\usepackage{bm}% bold math
\usepackage{scicite}
\usepackage{lineno}
\usepackage{physics}
\usepackage{newtxmath}
\usepackage{upgreek}
\usepackage{gensymb}
\usepackage{units}
\usepackage{float}
\usepackage{multirow}
\usepackage[inkscapelatex=false]{svg}
\usepackage[colorlinks=true,linkcolor=blue,urlcolor=blue,citecolor=blue]{hyperref}
\usepackage[commentmarkup=uwave,deletedmarkup=sout]{changes}
\definechangesauthor[name=Malte Grosche, color=blue]{fmg}
\usepackage{gensymb}
\usepackage{mathtools}
\usepackage{caption}
\newenvironment{sciabstract}{%
\begin{quote} \bf}
{\end{quote}}

\title{\vspace{-1cm} Strange metallicity encompasses high magnetic field-induced superconductivity in UTe$_2$}

\author{T. I. Weinberger,$^{1\dagger}$ H. Chen,$^{1\dagger}$ Z. Wu,$^{1\dagger}$ M. Long,$^{1}$\\  A. Cabala,$^{2}$ Y. Skourski,$^{3}$ J. Sourd,$^{3}$ T.~Haidamak,$^{2}$\\ V. Sechovský,$^{2}$ M. Vali{\v{s}}ka,$^{2}$ F. M. Grosche,$^{1}$ A. G. Eaton$^{1\ast}$\\
\\
\normalsize{$^{1}$Cavendish Laboratory, University of Cambridge,}\\
\normalsize{JJ Thomson Avenue, Cambridge, CB3 0HE, UK}\\
\normalsize{$^{2}$Charles University, Faculty of Mathematics and Physics, Department of}\\
\normalsize{ Condensed Matter Physics, Ke Karlovu 5, Prague 2, 121 16, Czech Republic}\\
\normalsize{$^{3}$Hochfeld-Magnetlabor Dresden (HLD-EMFL), Helmholtz-Zentrum }\\ \normalsize{Dresden-Rossendorf, Dresden, 01328, Germany}\\
\\
\normalsize{$^\ast$Email: alex.eaton@phy.cam.ac.uk}
\\
\normalsize{$^\dagger$These authors contributed equally to this work.}
}

\date{\today}

\begin{document}

\baselineskip24pt

\maketitle

\clearpage
%\linenumbers
\begin{sciabstract}

A strange metallic state -- characterized by an electrical resistivity that rises linearly in temperature $T$ at a rate governed by Planckian energetic dissipation -- is present in a number of unconventional superconductors. Despite widespread investigative efforts, the microscopic properties of strange metallicity continue to defy elementary description. Here we investigate heavy fermion UTe$_2$, which hosts a suite of exotic spin-triplet superconducting phases, the most extreme of which resides in a narrow angular window of intense magnetic fields $>$~40~T. Through angle-dependent magnetotransport measurements in pulsed magnetic fields, we find that this superconductive phase emerges from a strange metallic state with Planckian $T$-linear resistivity confined to a small region of phase space where the field-induced superconductivity is strongest, suggesting a shared underlying mechanism. These findings reveal a novel setting for strange metallicity -- in contrast to the general case of singlet superconductivity emerging on the border of antiferromagnetism, here strange metallicity accompanies field-induced triplet superconductivity within a spin-polarized magnetic state. This result demonstrates the remarkable ubiquity of strange metallicity across diverse materials settings, and highlights UTe$_2$ as a unique platform for exploring the interplay between unconventional superconductivity and quantum criticality.

\end{sciabstract}

\clearpage

\section*{\label{sec:intro}Introduction}

Landau's Fermi-liquid theory~\cite{landau1956theory} constitutes one of the most successful theoretical advances of 20$^{\text{th}}$ century physics. It describes the microscopic electronic properties of metals in terms of a ground-state quasiparticle model, enabling physicists to understand the thermodynamic and transport properties of metals with extraordinary precision. The elegance of the Fermi-liquid theory lies in its ability to describe the quasiparticle properties of interacting electronic systems in a framework analogous to the much less complicated case of free electrons. This simplicity has made it a powerful pillar underpinning much of modern condensed matter physics research.

However, in recent years a growing number of materials have been discovered to exhibit properties that contravene the Fermi-liquid paradigm~\cite{nFLStewartRevModPhys.73.797,RevModPhys2007nFL}. A hallmark of these so-called \textit{strange} metals is an electrical resistivity $\rho$ that rises linearly in temperature $T$, i.e. $\Delta\rho \propto T$, as opposed to the quadratic form $\Delta\rho \propto T^2$ characteristic of electron-electron scattering in a Fermi-liquid~\cite{landau1956theory}. Furthermore, many of these materials exhibit a scattering rate $\nicefrac{1}{\tau}$ that goes as $\tau \sim \frac{\hslash}{k_{\text{B}}T}$, where $k_{\text{B}}$ is Boltzmann's constant and $\hslash$ the reduced Planck constant~\cite{bruin2013similarity,HartnollRevModPhys.94.041002}. Commonly termed the Planckian dissipation rate, because this rate depends on no scale other than temperature, this signifies the presence of scale-invariant dynamics, which has evoked numerous comparisons with models of quantum gravity and holographic duality~\cite{phillips2022stranger}. Often found near a quantum critical point (QCP), strange metallic behavior has been conjectured to manifest due to scale-invariant critical fluctuations emanating from a QCP~\cite{GleisPRL25}. These may introduce nonlocal long-range interactions that destroy quasiparticle coherence, thereby forming a highly entangled quantum critical fluid in which the fundamental concept of an electronic particle -- one `quantum' of charge -- no longer appears to be valid~\cite{patel2023universal,chen2023shot}.

Especially pronounced in cuprate superconductors~\cite{MartinPRB90,daou2009linear,cooper2009anomalous,legros2019universal}, the identification of a strange metal state by the observation of $\Delta \rho \propto T$ at low $T$ has been reported in a range of materials families including heavy fermion compounds~\cite{CeCu6PRL94,TrovarelliPRL2000}, ruthenates~\cite{grigera2001magnetic,bruin2013similarity}, pnictides~\cite{analytis2014transport} and moir\'e-engineered bilayer graphene~\cite{polshyn2019large,CaoPhysRevLett.124.076801,jaoui2022quantum}. In several of these cases, strange metallicity emerges in a temperature range immediately above the critical temperature of an unconventional superconductive phase, hinting at a profound connection between these two phenomena. Moreover, the diversity of materials systems and experimental conditions in which strange metallic behavior has been observed points to some deep underlying principle of quantum physics that has yet to be fully uncovered.

Here we discover a new setting for Planckian strange metallicity. In high magnetic fields $\mu_0H >$~40~T the heavy fermion superconductor UTe$_2$ possesses an exotic field-induced superconductive state~\cite{Ranfieldboostednatphys2019}. By measuring the angular dependence of $\rho(T,H)$, we mapped the resistivity profile of UTe$_2$ in high magnetic fields. We find that in the region of phase space where the field-induced superconductivity is strongest, at $T > T_c$ strange metallicity is identified by the observation of $\Delta\rho \propto T$, which rises at the Planckian scattering rate. Our results add high magnetic field-induced, putatively spin-triplet superconductivity to the growing number of condensed matter settings in which strange metallicity is manifested.

\begin{figure}[t!]
\vspace{-2.1cm}
\begin{center}
\includegraphics[width=.9\linewidth]{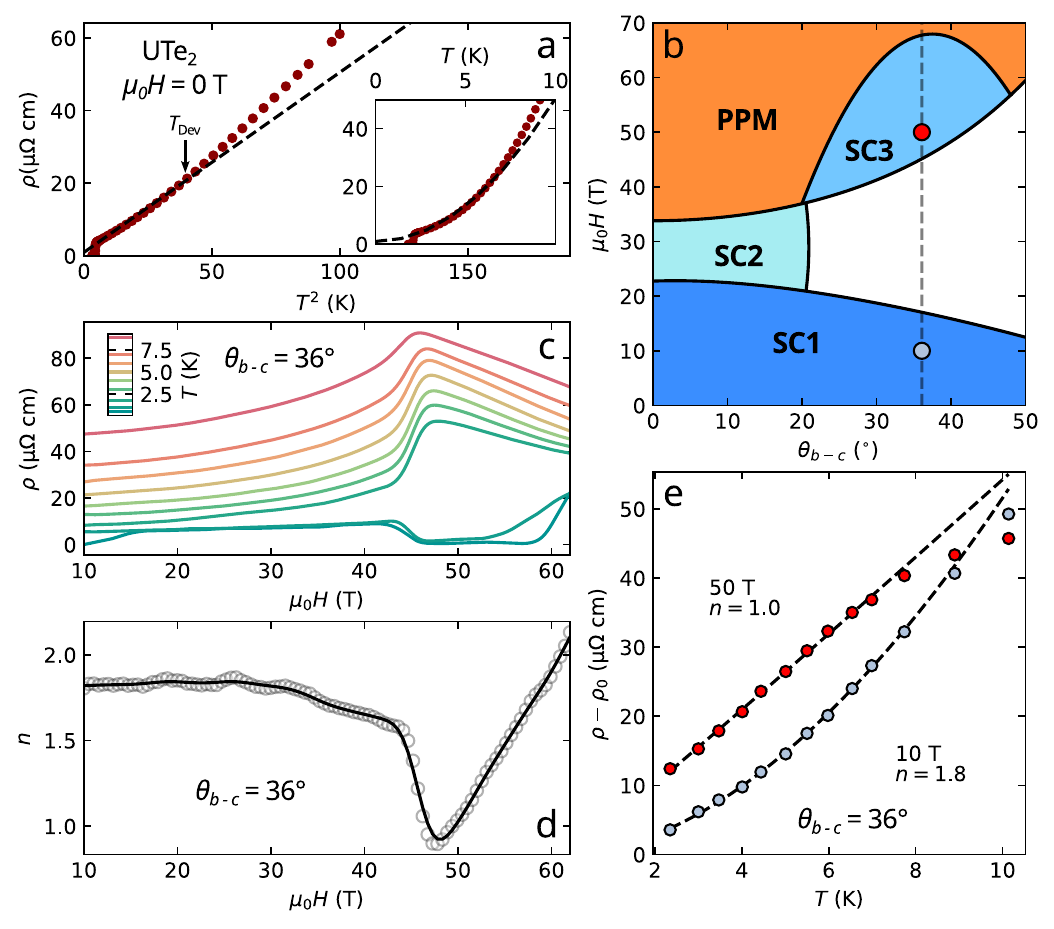}
\end{center}
\caption{\textbf{Magnetic field--induced $T$-linear resistivity. a,} Electrical resistivity $\rho$ of UTe$_2$ for current sourced along the [100] direction in zero applied magnetic field $H$ plotted against the square of the temperature $T$, with the inset plotted linearly in $T$. $\Delta \rho \propto T^2$ is observed at low temperatures, before the superconducting transition to $\rho = 0$ below $T_c$~=~2.1~K. Deviation from Fermi-liquid behavior is seen around $T_{\mathrm{Dev}} \approx$~6~K, above which the resistivity increases faster than quadratically. \textbf{b,} Schematic low temperature phase diagram of UTe$_2$ for $H$ versus angle $\theta_{b-c}$, defined as the rotation angle of magnetic field orientation away from the $b$-axis towards the $c$-axis. Three distinct superconducting (SC) phases are numbered~\cite{Ranfieldboostednatphys2019,tony2024enhanced}. The orange region represents a polarized paramagnetic (PPM) state. \textbf{c,} Selected isothermal pulsed field $\rho(H)$ curves between 0.5~K and 8.9~K for $H$ oriented at $\theta_{b-c}$~=~36$\degree$ (see Supplementary Materials for additional data). At 0.5~K zero resistance is observed at high $H$ as the SC3 state is accessed. At higher temperatures a sharp peak in $\rho(H)$ is observed at $\mu_0H \approx$~45~T, where the PPM state is entered upon crossing a first-order metamagnetic phase boundary~\cite{Aoki_UTe2review2022,lewin2023review}. \textbf{d,} The resistivity exponent $n$ from Eqn.~\ref{eq:exponent} as a function of $H$. This plot was constructed by binning the data into 100 equal-length intervals in the field range of 10~T to 62~T over the temperature range $T_c < T <$~7.5~K. The black solid line is a local spline fit giving a guide to the eye. \textbf{e,} Cuts in $T$ of the magnetotransport data at fixed field values of 10~T (grey) and 50~T (red). Fits to Eqn.~\ref{eq:exponent} are drawn as black dashed lines. At 10~T we obtain $n$~=~1.8, whereas at 50~T we find $n$~=~1.0.}
\label{fig:t-linear-BC36}
\end{figure}

\clearpage

\section*{\label{sec:intro}Results}
In zero applied field, the electrical resistivity of UTe$_2$ follows a $\rho=\rho_0+A T^2$ form at low $T$ down to the superconducting transition at $T_c$ = 2.1 K (Fig. \ref{fig:t-linear-BC36}a). The residual resistivity $\rho_0 \sim$ 1~$\upmu \Omega$~cm, documenting the long electronic mean free path of our samples.

Fig.~\ref{fig:t-linear-BC36}b sketches the low temperature phase diagram of UTe$_2$ for rotation of \textbf{H} from the crystalline $b$-axis ($\theta_{b-c} = 0\degree$) towards the $c$-axis~\cite{Ranfieldboostednatphys2019,tony2024enhanced}. Three separate superconducting phases are identified~\cite{lewin2023review}, the focus of the present study being the highest field SC3 state. The orange region of the phase diagram indicates a polarized paramagnetic state that is accessed for $\mu_0 H \geq$~35~T, at which point the magnetization suddenly jumps by approximately half a Bohr magneton per unit cell, characterizing a first-order metamagnetic transition~\cite{Miyake2019}. The metamagnetic transition field $H^*$ rises as \textbf{H} is tilted away from the $b$-axis. The SC3 phase occupies an angular range within the polarized paramagnetic state, which spans $\approx 25\degree$ at low $T$~\cite{helm2024,tony2025brief,qcl,lewin2025halo}. Prior studies of the SC3 state have reported its maximal critical temperature and field strength to be $T_c^{SC3} \approx$~2.4~K and $\mu_0H_{c2}^{SC3} \approx$~73~T~\cite{helm2024,tony2025brief} for \textbf{H} tilted to $\theta_{b-c} \approx$~35$\degree$.

In Fig.~\ref{fig:t-linear-BC36}c we plot the electrical resistivity of UTe$_2$ at incremental temperatures measured in pulsed magnetic fields by the standard four-terminal technique (see the Supplementary Materials for further details) with \textbf{H} orientated at $\theta_{b-c} =36\degree$. We extracted the temperature exponent $n$ of the resistivity by taking slices of the data at constant $H$, which were fitted to

\begin{align}
    \rho(T) = AT^n + \rho_0
\label{eq:exponent}
\end{align}

\noindent
where $A$ is a constant and $\rho_0$ the residual resistivity. Throughout this study we restrict our temperature domain for extracting $n$ to $T_c < T <$~7.5~K, over which range we find good power-law fitting with low residuals. By numerically fitting our $\rho(T,H)$ data to Eq.~\ref{eq:exponent}, we can therefore determine the evolution of $n(H)$.

\begin{figure}[t!]
\vspace{-0cm}
\begin{center}
\includegraphics[width=1\linewidth]{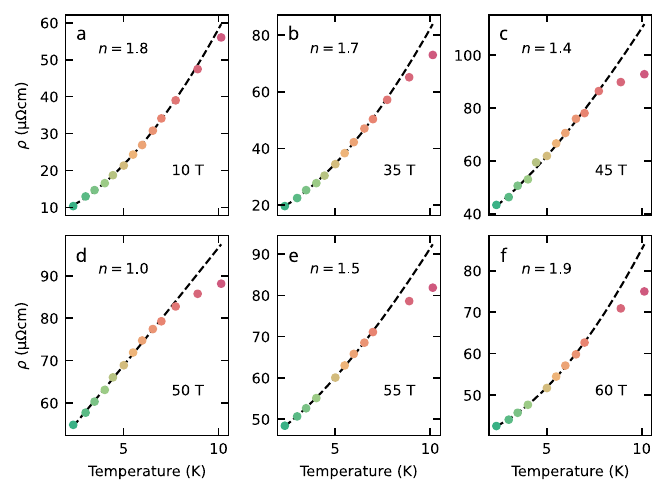}
\end{center}
\caption{\textbf{Evolution of the resistivity temperature exponent $n$ with magnetic field at $\theta_{b-c}$~=~36$\degree$. a-f,} Resistivity versus temperature at selected magnetic field cuts from 10~T up to 60~T, as indicated. For this orientation of \textbf{H}, the metamagnetic transition field $\mu_0H^* \approx$~45~T. For $H$ far below and above $H^*$, $n$ is close to 2. However, for $\mu_0H =$~50~T (panel d) $\rho(T)$ is linear at low $T$.}
\label{fig:t-linear-slices}
\end{figure}

At the modest magnetic field strength of 10~T, we find that $\rho(T)$ is close to the quadratic expectation of Fermi-liquid theory, with $n = 1.9$ (Fig.~\ref{fig:t-linear-BC36}c). By contrast, at 50~T -- having crossed $H^*$ into the polarized paramagnetic state -- the form of $\rho(T)$ has become linear, corresponding to $n$~=~1.0. In Fig.~\ref{fig:t-linear-BC36}d we plot the evolution of $n(H)$, which exhibits a sharp drop to $n =$~1 upon crossing $H^*$ (that for this angle is at around 45~T), before rising back up towards $n=$~2 when $\mu_0H =$~60~T. In Figure~\ref{fig:t-linear-slices} we re-plot the data of Fig.~\ref{fig:t-linear-BC36} as a function of $T$, by taking slices at fixed values of $H$. These $\rho(T)$ plots at constant $H$ confirm the non-monotonic field-dependence of the power-law exponent $n$ shown in Fig.~\ref{fig:t-linear-BC36}d.

\begin{figure}[p]
\vspace{-1cm}
\begin{center}
\includegraphics[width=.8\linewidth]{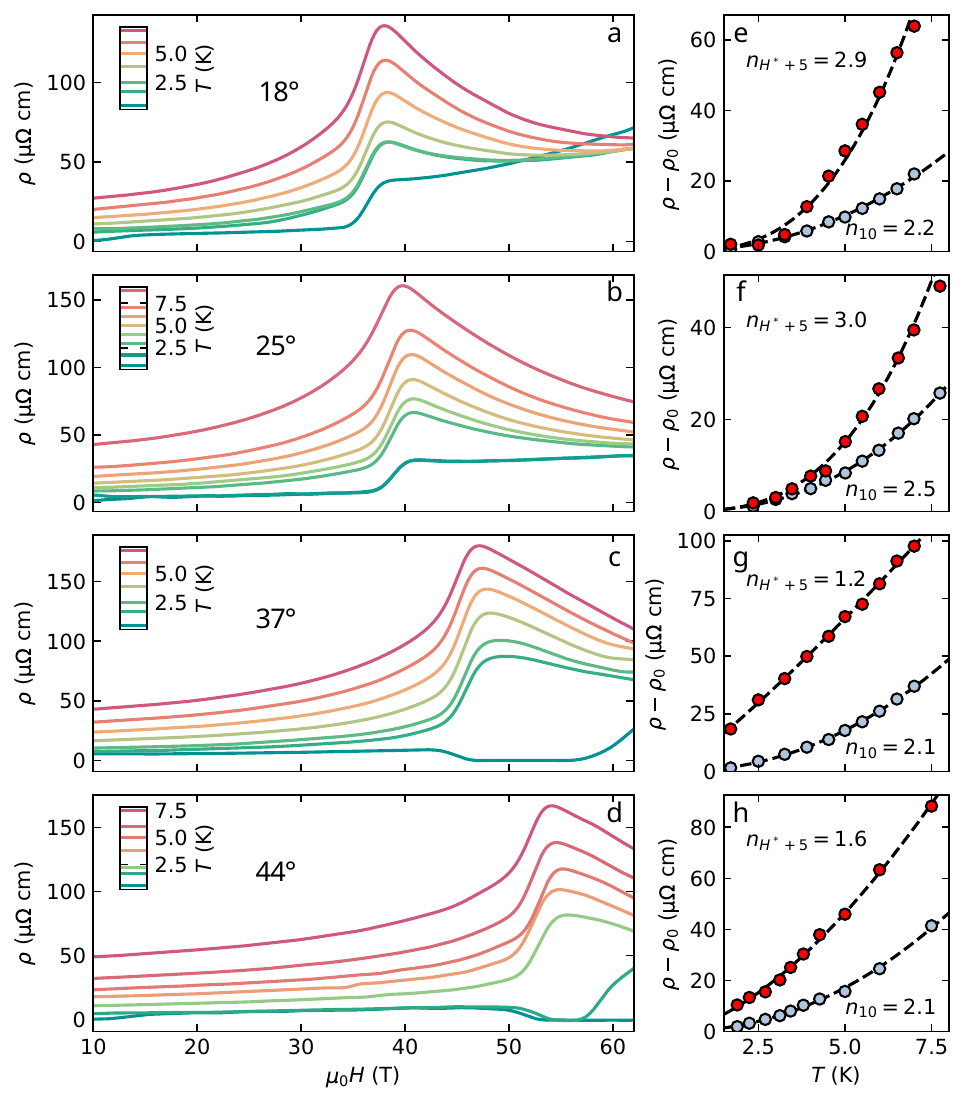}
\end{center}
\caption{\textbf{Angular-dependent magnetotransport measurements. a-d} Isothermal $\rho(H)$ curves at indicated temperatures and inclinations of $\theta_{b-c}$. Only selected curves are plotted, for clarity of presentation -- see Supplementary Materials for the full dataset. \textbf{e-h} Cuts of data from panels (a-d) at field values of $\mu_0H =$~10~T (grey points, $n_{10}$) and at a fixed 5~T above the metamagnetic transition field $H^*$ (red points, $n_{H^* + 5}$) for each angle. While $n_{10}$ exhibits only a small variation over this range, always being close to quadratic, by contrast the variation in $n_{H^* + 5}$ is over three times larger, ranging in profile from being close to cubic at $\theta_{b-c} = 25\degree$ to close to linearity at $\theta_{b-c} = 37\degree$.}
\label{fig:t-linear-angles}
\end{figure}

% We note that at $H=0$, $\rho(T)$ deviates from the low-$T$  quadratic profile for $T \gtrapprox 6$~K (Fig.~\ref{fig:t-linear-BC36}a). This behavior has been noted by other UTe$_2$ studies, and discussed in the context of signatures in the magnetization, electronic specific heat and thermal expansion that appear to indicate the presence of some form of Kondo or magnetic coherence occurring at $T \approx$~12~K~\cite{Aoki_UTe2review2022,Willa104.205107}. Therefore, throughout the present study, we restrict our temperature domain for extracting $n$ to $T_c < T <$~7.5~K, over which range we find good power-law fitting with low residuals.

The profile of $n(H)$ in UTe$_2$ depends strongly on the orientation of \textbf{H}. Figure~\ref{fig:t-linear-angles} presents raw $\rho(H)$ traces at incremental temperatures for four different angles in the range $18\degree \leq \theta_{b-c} \leq 44\degree$. We also plot ($\rho - \rho_0$) as a function of $T$ for each orientation at low field (grey points) and high field (red points). As the metamagnetic transition field $H^*$ increases with increasing rotation angle away from the $b$-axis approximately following a $\nicefrac{1}{\cos{\theta_{b-c}}}$ dependence~\cite{helm2024}, the red points are taken at a fixed interval of 5~T above $\mu_0 H^*$, to follow how the electrical transport evolves in proximity to the upper boundary of the transition. The grey points are each taken at a fixed value of $\mu_0H =$~10~T. At 10~T there is little variation in $n(\theta_{b-c})$, which sits in the range 2.1~$\leq n \leq$~2.5 over this angular interval. By contrast, at 5~T above $\mu_0 H^*$ there is a large spread in $n(\theta_{b-c})$ from 3.0 at $\theta_{b-c} = 25\degree$ down to 1.0 at $\theta_{b-c} = 36\degree$. The regions of super-quadratic $\rho(T)$ indicate the presence of spin-wave scattering~\cite{kubo1972quantum}, while the sub-quadratic regime points to the breakdown of Fermi-liquid phenomenology.

\begin{figure}[p]
\vspace{-0cm}
\begin{center}
\includegraphics[width=0.8\linewidth]{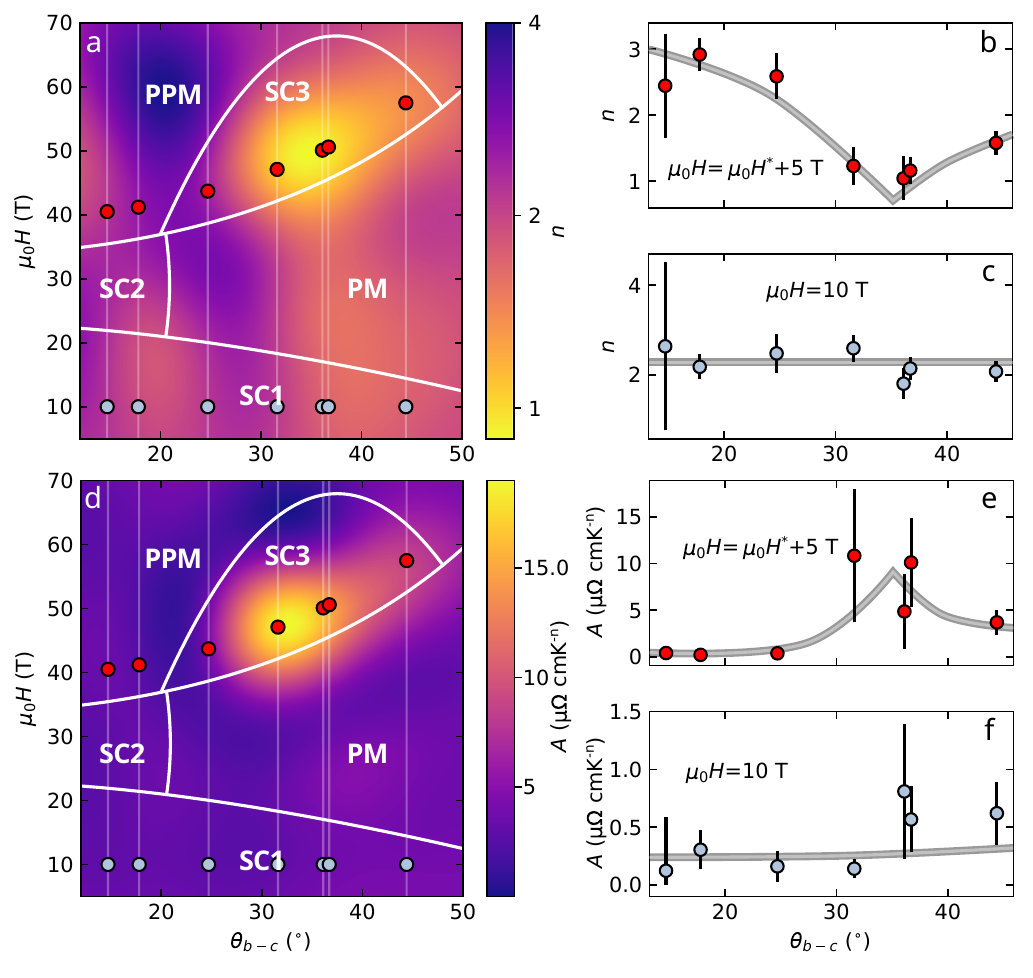}
\end{center}
\caption{\textbf{Strange metallicity underpins magnetic field--induced superconductivity in UTe$_2$.  a,} Heatmap of the resistivity temperature exponent $n$ from Eqn.~\ref{eq:exponent}. PM denotes the normal paramagnetic state. The bright yellow region, corresponding to $n = 1$, is located at the centre of the SC3 dome. \textbf{b,} Evolution of $n$ with $\theta_{b-c}$ at a fixed value of 5~T above the metamagnetic transition field $\mu_0H^*$, which rises as $\nicefrac{1}{\cos{\theta_{b-c}}}$. These points are marked in red in panel (a). For small $\theta_{b-c}$, $2 < n < 3$ for these field values. Upon rotating the orientation of \textbf{H} to higher $\theta_{b-c}$, $n$ decreases to $n = 1.0$ at $\theta_{b-c} = 36\degree$, before rising again under further rotation. \textbf{c,} By contrast, at a fixed field value of $\mu_0H =$~10~T, $n =$~2 (within error) over the whole measured interval of $\theta_{b-c}$. The procedure for calculating error bars is detailed in the Supplementary Materials. \textbf{d,} Heatmap of the resistivity $A$ coefficient from Eqn.~\ref{eq:exponent}. \textbf{e,} The evolution of $A$ at 5~T above $\mu_0H^*$ and \textbf{f,} at $\mu_0H =$~10~T. Note that as $n$ changes, so do the dimensions of $A$, giving a greater relative error bar (see Supplementary Materials for details). A pronounced peak in $A(H,\theta_{b-c})$ is observed in the strange metallic state, over an order of magnitude greater than at low field.}
\label{fig:t-linear-heat}
\end{figure}

In Figure~\ref{fig:t-linear-heat} we track the angular evolution of $n(\theta_{b-c})$ with a heatmap. Bright yellow coloring corresponds to low values of $n$ close to 1, with darker colors representing higher values of $n$. We also plot the evolution of $A(\theta_{b-c})$ with a heatmap. Because the dimensions of $A$ depend on the value of the exponent $n$, here we have taken the initial increment of the fitted resistivity between 0 K and 1 K for each angle, which thereby illustrates the strength of electronic scattering at low temperatures. The most striking aspect of Fig.~\ref{fig:t-linear-heat} is the close correspondence between the region of $T$-linear resistivity -- exhibiting high values of $A$ an order of magnitude greater than at low field -- which closely coincides with the angular domain of the SC3 phase. There is a large section of bright yellow coloring, indicative of strange metallicity, which is centred around $\theta_{b-c} \approx 35 \degree$. This orientation is where prior studies have identified that the SC3 state extends to its highest critical field strength and temperature~\cite{helm2024,tony2025brief}. Our experiments therefore reveal a strong connection between strange metallicity and SC3 superconductivity in a tightly confined pocket of the high-field phase diagram.

In the Supplementary Materials we consider the quasiparticle scattering rate implied by the linear slope of $\rho(T)$ for the high field region of the phase diagram where $\Delta \rho \propto T$. We performed density functional theory calculations based on the well-characterized low field Fermi surface of UTe$_2$~\cite{WrayARPES_PhysRevLett.124.076401,AokidHvA_UTe2-2022,Eaton2024,theo2024,weinberger2025pressure} to estimate $\nicefrac{1}{\tau}$ from the bare and renormalized plasma frequencies. We complemented this analysis with fittings of the SC3 domain along with a consideration of the implied coherence length of SC3. In combination, these analyses let us estimate a range for the dimensionless parameter $\alpha$ defined as $\alpha \equiv \frac{\hbar}{k_{\text{B}}T \tau}$. We find that $\alpha = 1.3 \pm 0.6$, indicating that the electronic scattering occurs at the Planckian rate (defined as $\alpha$ equal to unity, see Supplementary Materials for detailed calculations). Our study therefore adds high magnetic field UTe$_2$ to the growing list of Planckian strange metals~\cite{bruin2013similarity,HartnollRevModPhys.94.041002,phillips2022stranger}.

%Since the dimensions of $A$ depend on the value of the exponent $n$, here we  $\Delta\rho(1 \mathrm{K}) = \rho(1 \mathrm{K}) - \rho_0$ from our fit to the data, as a measure of the low temperature quasiparticle scattering and mass. We plot the evolution of $\Delta\rho(1 \mathrm{K})(\theta_{b-c})$ with a heatmap, and the change of both $n$ and $\Delta\rho(1 \mathrm{K})$ at the fixed field values of 10~T and at 5~T above $\mu_0 H^*$. The most striking aspect of Fig.~\ref{fig:t-linear-heat} is the close correspondence between the region of $T$-linear resistivity -- exhibiting high values of $\Delta\rho(1 \mathrm{K})$ an order of magnitude greater than at low field -- which closely coincides with the angular domain of the SC3 phase. There is a large section of bright yellow coloring, indicative of strange metallicity, which is centred around $\theta_{b-c} \approx 35 \degree$. This orientation is where prior studies have identified that the SC3 state extends to its highest critical field strength and temperature~\cite{helm2024,tony2025brief}. Our experiments therefore reveal a strong connection between the magnetic field--inducement of both strange metallicity and superconductivity in UTe$_2$ for a narrow angular range of \textbf{H}.

%\clearpage

\section*{\label{sec:intro}Discussion}

Several heavy fermion compounds have been found to exhibit non-Fermi-liquid transport properties in proximity to quantum criticality, typically in the vicinity of an antiferromagnetic QCP~\cite{TrovarelliPRL2000,PaglionePhysRevLett.91.246405,BianchiPhysRevLett.91.257001,steppke2013ferromagnetic,shen2020CeRh6Ge4}. Ferromagnetic phases tend not to terminate at a QCP, either due to the imposition of a first-order transition, or because another phase of e.g. helical or spin-density-wave ordering intervenes before the Curie temperature can be tuned to zero~\cite{BrandoRevModPhys.88.025006,friedemann2018quantum}. Notable exceptions include As-doped YbNi$_4$P$_2$~\cite{steppke2013ferromagnetic} and compressed CeRh$_6$Ge$_4$~\cite{shen2020CeRh6Ge4}.

Our findings here, revealing the presence of non-Fermi-liquid normal state transport properties over a narrow angular range close to $\theta_{b-c} \approx 35\degree$, strongly point to the presence of quantum criticality. Yet, intriguingly, the SC3 state has been extensively studied by several groups~\cite{Ranfieldboostednatphys2019,knafo2021comparison,lewin2023review,Miyake2021,LANL_bulk_UTe2,helm2024,tony2025brief,qcl,lewin2025halo} and except for metamagnetism no other thermodynamic, magnetic, or structural anomaly has been discerned near $\theta_{b-c} \approx 35\degree$. This raises the question: is there another form of order that reaches quantum criticality within the SC3 region?

 For $H>H^*$, the polarized paramagnetic state of UTe$_2$ is characterized by localized magnetic moments of 0.5~$\mu_{\text{B}}$ per uranium ion~\cite{Aoki_UTe2review2022}, which is qualitatively similar to a ferromagnetic state. In our prior work we found that the critical end point of the first-order metamagnetic transition into the spin-polarized state can be continuously tuned to zero temperature -- a quantum critical end point -- by tilting the orientation of \textbf{H} in any direction away from the hard magnetic $b$-axis~\cite{qcl}. This traces out a quantum critical line (QCL) in the high field phase diagram, to which the superconducting SC3 phase appears to be anchored. However, SC3 lies inside the QCL rather than being centred around it, as might be expected if the QCL were the sole pairing-mediation source. Whereas the quantum critical boundary of the metamagnetic transition surface is reached for comparatively moderate transverse fields in the $a$-direction, much larger $c$-axis fields are required to reach the quantum critical boundary: the metamagnetic transition stays sharp and well resolved up to at least $\theta_{b-c} = 60\degree$~\cite{qcl}. This strongly suggests that critical metamagnetic fluctuations associated with the QCL are not responsible for the linear temperature dependence of the resistivity we observe here at $\theta_{b-c} \approx 35\degree$, where SC3 also happens to have its maximal $T_c$ and $H_{c2}$ (in the $b-c$ plane).

%A likely suspect for this parameter is a component of the magnetic susceptibility tensor $\boldsymbol\chi$. In general $\boldsymbol\chi$ is a second-order tensor comprising nine components $\chi_{ij}$, where $i,j$ refer to the principal crystallographic axes~\cite{constable1990bootstrap}. It appears clear that the longitudinal $\chi_{bb}$ component does not diverge here, as the metamagnetic transition is still sharply first-order-like for 20$\degree < \theta_{b-c} <$~40$\degree$~\cite{qcl,Miyake2021}. However, one (or more) of the other components might be diverging in the strange metallic region. This may introduce additional complexity to the high-$H$ phase landscape of UTe$_2$~\cite{ripples}. A detailed knowledge of how the components of $\boldsymbol\chi$ evolve in high magnetic fields could inform microscopic models of the pairing interaction underpinning the formation of SC3, and by extension provide insight into the likely \textbf{d}-vector structure of the triplet order parameter symmetry~\cite{APMackenzieRevModPhys.75.657}.

There is strong evidence favoring the scenario of odd-parity pairing in UTe$_2$ at low $T$ for $\mu_0H$~=~0~T~\cite{lewin2023review,Aoki_UTe2review2022}. The likelihood for Cooper pairs to possess a (pseudo)spin-triplet state in the high-$H$ superconductive phases of UTe$_2$ is arguably even greater than it is at low $H$. This makes our observation of strange metallicity encompassing the SC3 region all the more curious, as although $T$-linear resistivity often accompanies even-parity superconductivity~\cite{phillips2022stranger}, it has not previously been associated with a spin-triplet superconductive state. Our discovery therefore poses fresh questions pertaining to the interplay between superconductive pairing and non-Fermi-liquid phenomenology.

Especially puzzling is the fact that in UTe$_2$ the strange metal state (and associated field-induced superconductivity) is so acutely sensitive to the inclination of \textbf{H}. The simplicity of the low $H$ Fermi surface~\cite{Eaton2024} makes a Lifshitz transition localized to $\theta_{b-c} \approx 35\degree$ unlikely. This angular dependence also appears difficult to reconcile within a more general scenario such as the paramagnetic destruction of Kondo singlets being the leading mechanism driving quasiparticle decoherence. Instead, it suggests some competition between magnetic interactions that possess similar (low) energy scales. Another layer of complexity lies in the fact that no phase boundary is so far known to be continuously suppressed over this angular interval~\cite{lewin2023review}. We speculate that our finding of strange metallicity encompassing SC3 points to the emergence of an additional, so far unidentified, order parameter. This could plausibly be some sort of spin density instability~\cite{ripples} analogous to the cases of Sr$_3$Ru$_2$O$_7$~\cite{lester2015field,lester2021magnetic} and URu$_2$Si$_2$~\cite{knafo2016field} -- or perhaps a new form of hidden order. Bulk-sensitive thermodynamic measurements across the high field strange metal region of UTe$_2$ would help to shed light on the remarkably strange physics manifested by this material.

%In conclusion, we measured the electrical resistivity of UTe$_2$ in high magnetic fields. In the region of the magnetic field strength--tilt angle phase diagram where magnetic field-induced superconductivity is observed up to $\gtrapprox$~70~T at low temperatures, at elevated temperatures the normal state resistivity exhibits characteristic features of a strange metallic state, with a greatly enhanced $\Delta\rho(1 \mathrm{K})$ coefficient and a temperature exponent $n = 1.0$. The strange metal region of the phase diagram closely overlaps with the domain of high field-induced superconductivity, strongly suggesting an intimate connection between these two phenomena.

%chat about triplet SC, can't work out planckian, 'entanglement' nonsense etc

\clearpage
\normalsize{
\bibliographystyle{Science}
\bibliography{UTe2}
}

\clearpage
\Large
\vspace{12mm}\noindent
\textbf{Acknowledgements}

\normalsize
\vspace{5mm}\noindent 
We gratefully acknowledge stimulating discussions with D. Chichinadze, D. Shaffer, J. Yu, S. Raghu, and especially G. Lonzarich. This project was supported by the EPSRC of the UK through grants EP/X011992/1 EP/Z533695/1 and EP/R513180/1. We acknowledge support of the HLD at HZDR, a member of the European Magnetic Field Laboratory (EMFL) and from EPSRC UK via its membership to the EMFL (grant EP/N01085X/1). Crystal growth and characterization were performed in MGML (mgml.eu), which is supported within the program of Czech Research Infrastructures (project no. LM2023065). We acknowledge financial support by the Czech Science Foundation GAČR under the Junior Star Grant No. 26-21795M (STiUS). T.I.W. acknowledges support from Murray Edwards College (University of Cambridge) and the Cambridge Philosophical Society through a Henslow Fellowship. A.G.E. acknowledges support from Sidney Sussex College (University of Cambridge).

%\clearpage

\end{document}